\documentclass[runningheads]{llncs}

\usepackage[T1]{fontenc}
\usepackage{graphicx}
\usepackage{xcolor, colortbl}
\usepackage{amsmath}
\usepackage{amssymb}
\usepackage{rotating}
\usepackage{semantic}
\usepackage{wasysym}
\usepackage{changepage}
\usepackage{tabularx}

\makeatletter
\DeclareSymbolFont{extraup}{U}{zavm}{m}{n}
\DeclareMathSymbol{\newcheckmark}{\mathalpha}{extraup}{128}
\DeclareMathSymbol{\newcrossmark}{\mathalpha}{extraup}{129}
\makeatother

\definecolor{highlight}{rgb}{0.96,0.96,0.96}

\newcommand{\ctr}[1]{\textsc{\small #1}}
\newcommand{\fun}[1]{\textsf{#1}}
\newcommand{\key}{\textbf}
\newcommand{\typ}{\textit}
\newcommand{\hole}[1]{\CIRCLE_{#1}}

\newcommand{\myth}{\textsc{Myth}}
\newcommand{\smyth}{\textsc{Smyth}}
\newcommand{\scrybe}{\textsc{Scrybe}}
\newcommand{\lam}{$\lambda^2$}
\newcommand{\betaeta}{\textit{$\beta$-normal, $\eta$-long}}
\newcommand{\yes}{$\newcheckmark$}
\newcommand{\no}{$\newcrossmark$}
\newcommand{\total}{$^\dagger$}
\newcommand{\indexed}{$^\ddagger$}
\newcommand{\constraint}[1]{
  \left\{
  \begin{array}{c}
    #1
  \end{array}
  \right\}
}

\newcommand\doubleplus{\ensuremath{\mathbin{+\mkern-10mu+}}}

\begin{document}

\title{Program Synthesis Using Example Propagation}



\author{Niek Mulleners\inst{1}\orcidID{0000-0002-7934-6834} \and
Johan Jeuring\inst{1}\orcidID{0000-0001-5645-7681} \and
Bastiaan Heeren\inst{2}\orcidID{0000-0001-6647-6130}}

\authorrunning{N. Mulleners et al.}

\institute{Utrecht University, The Netherlands,
\email{\{n.mulleners,j.t.jeuring\}@uu.nl} \and Open University of The
Netherlands, The Netherlands, \email{bastiaan.heeren@ou.nl}}

\maketitle

\begin{abstract}
  We present \scrybe{}, an example-based synthesis tool for a statically-typed
  functional programming language, which combines top-down deductive reasoning
  in the style of \lam{} with \smyth{}-style live bidirectional evaluation.
  During synthesis, example constraints are propagated through sketches to
  prune and guide the search. This enables \scrybe{} to make more effective use
  of functions provided in the context. To evaluate our tool, it is run on the
  combined, largely disjoint, benchmarks of \lam{} and \myth{}. \scrybe{} is
  able to synthesize most of the combined benchmark tasks.


\end{abstract}

\section{Introduction}


Type-and-example driven program synthesis is the process of automatically
generating a program that adheres to a type and a set of input-output examples.
The general idea is that the space of type-correct programs is enumerated,
evaluating each program against the input-output examples, until a program is
found that does not result in a counterexample. Recent work in this field has
aimed to make the enumeration of programs more efficient, using various pruning
techniques and other optimizations. \textsc{Hoogle+}~\cite{guo_2019} and
\textsc{Hectare}~\cite{koppel_2022} explore efficient data structures to
represent the search space. Smith and Albarghouthi~\cite{smith_2019} describe
how synthesis procedures can be adapted to only consider programs in normal
form. \textsc{MagicHaskeller}~\cite{katayama_2008} and RESL~\cite{peleg_2020}
filter out programs that evaluate to the same result.
Instead of only using input-output examples for the verification of generated
programs, \myth{}~\cite{osera_2015,osera_2016}, \smyth{}~\cite{lubin_2020}, and
\lam{}~\cite{feser_2015} use input-output examples during pruning, by
eagerly checking incomplete programs for counterexamples using constraint
propagation.

\subsubsection{Constraint Propagation}

Top-down synthesis incrementally builds up a sketch, a program which may
contain holes (denoted by $\hole{}$). Holes may be annotated with constraints,
e.g.~type constraints. During synthesis, holes are filled with new sketches
(possibly containing more holes) until no holes are left. For example, for
type-directed synthesis, let us start from a single hole $\hole{0}$ annotated
with a type constraint:
\[
  \hole{0} :: \typ{List}~ \typ{Nat} \rightarrow \typ{List}~ \typ{Nat}
\]
We may fill $\hole{0}$ using the function $\fun{map}::(a \rightarrow b)
\rightarrow\typ{List}~a\rightarrow\typ{List}~b$, which applies a function to
the elements of a list. This introduces a new hole $\hole{1}$, with a new type
constraint:
\[
  \hole{0} :: \typ{List}~ \typ{Nat} \rightarrow \typ{List}~ \typ{Nat}
  \quad
  \xrightarrow{\fun{map}}
  \quad
  \fun{map}~ (\hole{1} :: \typ{Nat} \rightarrow \typ{Nat})
\]
We say that the constraint on $\hole{0}$ is propagated through \fun{map} to the
hole $\hole{1}$. Note that type information is preserved: the type constraint
on $\hole{0}$ is satisfied exactly if the type constraint on $\hole{1}$ is
satisfied. We say that the hole filling \fbox{$\hole{0} \mapsto
\fun{map}~\hole{1}$} refines the sketch with regards to its type constraint.

A similar approach is possible for example constraints, which partially specify
the behavior of a function using input-output pairs. For example, we may
further specify hole $\hole{0}$, to try and synthesize a program that doubles
each value in a list:\footnote{In this example, as well as in the rest of this
paper, we will leave type constraints implicit.}
\[
  \hole{0} \vDash \constraint{[0,1,2] \mapsto [0,2,4]}
\]
Now, when introducing \fun{map}, we expect its argument $\hole{1}$ to have
three example constraints, representing the doubling of a natural number:
\[
  \hole{0} \vDash \constraint{[0,1,2] \mapsto [0,2,4]}
  \quad
  \xrightarrow{\fun{map}}
  \quad
  \fun{map}~ (\hole{1} \vDash
    \left\{
    \begin{array}{c}
      0 \mapsto 0 \\
      1 \mapsto 2 \\
      2 \mapsto 4
    \end{array}
    \right\}
    )
\]
Similar to type constraints, we want example constraints to be correctly
propagated through each hole filling, such that example information is
preserved. Unlike with type constraints, which are propagated through
hole fillings using type checking/inference, it is not obvious how to propagate
example constraints through arbitrary functions. Typically, synthesizers define
propagation of example constraints for a hand-picked set of functions and
language constructs. Feser et al.~\cite{feser_2015} define example propagation
for a set of combinators, including \fun{map} and \fun{foldr}, for their
synthesizer~\lam{}. Limited to this set of combinators, \lam{} excels at
composition, but lacks in generality. \myth{}~\cite{osera_2015,osera_2016}, and
by extension \smyth{}~\cite{lubin_2020}, take a more general approach, in
exchange for compositionality, defining example propagation for basic language
constructs, including constructors and pattern matches.

\begin{figure}[t]
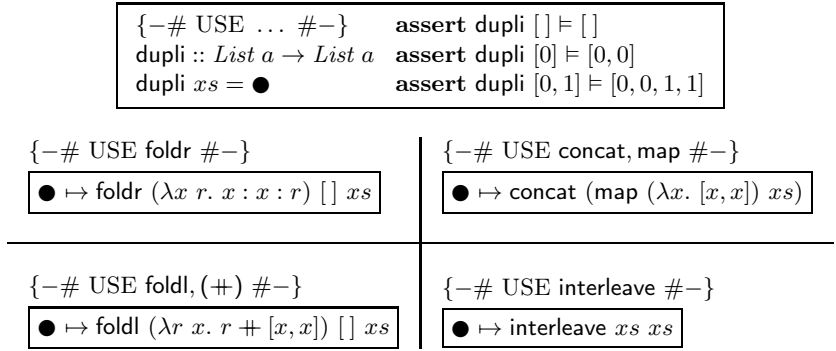

\[
  \begin{array}{c}
    \boxed{
    \begin{array}{ll}
      \{-\#~ \text{USE}~ \dots ~\#-\} &
      \key{assert}~ \fun{dupli}~ [\:] \vDash [\:] \\
      \fun{dupli} :: \typ{List}~ a \rightarrow \typ{List}~ a &
      \key{assert}~ \fun{dupli}~ [0] \vDash [0,0] \\
      \fun{dupli}~ xs = \hole{} &
      \key{assert}~ \fun{dupli}~ [0,1] \vDash [0,0,1,1] \\
    \end{array}
    } \\ \\
    \begin{array}{l|l}
      \begin{array}{l}
        \{-\#~ \text{USE}~ \fun{foldr} ~\#-\} \\[2pt]
        \boxed{\hole{} \mapsto \fun{foldr}~ (\lambda x~ r .~ x:x:r)~ [\:]~ xs}
      \end{array}
      &
      \begin{array}{l}
        \{-\#~ \text{USE}~ \fun{concat}, \fun{map} ~\#-\} \\[2pt]
        \boxed{\hole{} \mapsto \fun{concat}~ (\fun{map}~ (\lambda x .~ [x,x])~ xs)}
      \end{array}
      \\ \\
      \hline
      \\
      \begin{array}{l}
        \{-\#~ \text{USE}~ \fun{foldl}, \fun{(\doubleplus)} ~\#-\} \\[2pt]
        \boxed{\hole{} \mapsto \fun{foldl}~ (\lambda r~ x.~ r \doubleplus [x,x])~ [\:]~ xs}
      \end{array}
      &
      \begin{array}{l}
        \{-\#~ \text{USE}~ \fun{interleave} ~\#-\} \\[2pt]
        \boxed{\hole{} \mapsto \fun{interleave}~ xs~ xs}
      \end{array}
    \end{array}
  \end{array}
\]
\caption{(Top) A program sketch in \scrybe{}, for synthesizing the function
  \fun{dupli}, which duplicates each value in a list. (Bottom) Different
  synthesis results returned by \scrybe{}, for different sets of included
  functions.}
\label{fig:dupli}
\end{figure}

\subsubsection*{Presenting \scrybe{}}
In this paper, we explore how the techniques of \lam{} and \smyth{} can be
combined to create a general-purpose, compositional example driven synthesizer,
which we will call \scrybe{}. Figure~\ref{fig:dupli} shows four different
interactions with \scrybe{}, where the function \fun{dupli} is synthesized with
different sets of functions. \scrybe{} is able to propagate examples through
all of the provided functions using live bidirectional evaluation as introduced
by Lubin et al.~\cite{lubin_2020} for their synthesizer \smyth{}, originally
intended to support sketching~\cite{solar-lezama_2008,solar-lezama_2009}. By
choosing the right set of functions (for example, the set of combinators used
in \lam{}), \scrybe{} is able to cover different synthesis domains.
Additionally, allowing the programmer to choose this set of functions opens up
a new way for them to express their intent to the synthesizer, without going
out of their way to provide an exact specification.



\subsubsection*{Main Contributions}
The contributions of this paper are as follows:

\begin{itemize}
  \item We give an overview of example propagation and how it can be used to
    perform program synthesis (Section~\ref{sec:example_propagation}).
  \item We show how live bidirectional evaluation as introduced by Lubin et
    al.~\cite{lubin_2020} allows arbitrary sets of functions to be used as
    refinements during program synthesis (Section~\ref{sec:techniques}).
  \item We present \scrybe{}, an extension of \smyth{}~\cite{lubin_2020} and
    evaluate it against existing benchmarks from different synthesis domains
    (Section~\ref{sec:benchmark}).
\end{itemize}


\section{Example Propagation}\label{sec:example_propagation}




Example constraints give a specification of a function in terms of input-output
pairs. For example, the following constraint represents the function \fun{mult}
that multiplies two numbers.
\[
  \left\{
  \begin{array}{c}
    0 \ \ 1 \mapsto 0 \\
    1 \ \ 1 \mapsto 1 \\
    2 \ \ 3 \mapsto 6
  \end{array}
  \right\}
\]
The constraint consists of three input-output examples. Each arrow ($\mapsto$)
maps the inputs on its left to the output on its right. A function can be
checked against an example constraint by evaluating it on the inputs and
matching the results against the corresponding outputs. During synthesis, we
want to check that generated expressions adhere to these constraints. For
example, to synthesize \fun{mult}, we may generate a range of expressions of
type $\typ{Nat} \rightarrow \typ{Nat} \rightarrow \typ{Nat}$ and then check
each against the example constraint. The expression $\lambda x~ y .~
\fun{double}~ (\fun{plus}~ x~ y)$ will be discarded, as it maps the inputs to
$2$, $4$ and $10$, respectively. It would be more efficient, however, to
recognize that any expression of the form $\lambda x~ y .~ \fun{double}~ e$,
for some expression $e$, can be discarded, since there is no natural number
whose double is~$1$.

To discard incorrect expressions as early as possible, we incrementally
construct a sketch, where each hole (denoted by $\hole{}$) is annotated with an
example constraint. Each time a hole is filled, the example constraints are
propagated to the new holes and checked for contradictions. Let us start from a
single hole $\hole{0}$. We refine the sketch by eta-expansion, binding the
inputs to the variables $x$ and~$y$.
\[
  \hole{0} \vDash
  \left\{
  \begin{array}{c}
    0 \ \ 1 \mapsto 0 \\
    1 \ \ 1 \mapsto 1 \\
    2 \ \ 3 \mapsto 6
  \end{array}
  \right\}
  \quad
  \xrightarrow{\textbf{eta-expand}}
  \quad
  \lambda x~ y .~ (\hole{1} \vDash
  \left\{
  \begin{array}{cc|c}
    x & y \\
    0 & 1 & 0 \\
    1 & 1 & 1 \\
    2 & 3 & 6
  \end{array}
  \right\})
\]
A new hole $\hole{1}$ is introduced, annotated with a constraint that captures
the values of $x$ and $y$. Example propagation through \fun{double} should be
able to recognize that the value $1$ is not in the codomain of \fun{double}, so
that the hole filling $\boxed{\hole{1} \mapsto \fun{double}~ \hole{2}}$ can be
discarded.

\subsection{Program Synthesis Using Example Propagation}

Program synthesizers based on example propagation iteratively build a program
by filling holes. At each iteration, the synthesizer may choose to fill a hole
using either a \textit{refinement} or a \textit{guess}. A \textit{refinement}
is an expression for which example propagation is defined. For example,
eta-expansion is a refinement, as shown in the previous example. To propagate
an example constraint through a lambda abstraction, we simply bind the inputs
to the newly introduced variables. A \textit{guess} is an expression for which
example propagation is \emph{not} defined. The new holes introduced by a guess
will not have example constraints. Once you start guessing, you have to keep
guessing! Only when all holes introduced by guessing are filled can the
expression be checked against the example constraint. In a sense, guessing
comes down to brute-force enumerative search.

\noindent
Refinements are preferred over guesses, since they preserve constraint
information, which is needed to prune the search space. It is, however, not
feasible to define example propagation for every possible expression. Instead,
previous synthesizers define only a hand-picked set of refinements. In the rest
of this section, we show how the synthesizers \lam{}, \myth{} and \smyth{}
implement and use example propagation.



\subsection{Example Propagation in \lam{}}


For the tool \lam{}, Feser et al.~\cite{feser_2015} define deduction rules for
a set of combinators, including \fun{map}, \fun{foldr}, and \fun{filter}. In
essence, these deduction rules propagate examples through the respective
combinators. For example, consider \fun{map}, which maps a function over a
list. Refinement using \fun{map} replaces a constraint on a list with
constraints on its elements, while checking that the input and output lists
have equal length, and that no value in the input list is mapped to different
values in the output list.

\begin{figure}[t]
\[
  \begin{array}{rccc}
    & \fun{inc} & \fun{compress} & \fun{reverse} \\[2pt]
    \hole{0} \vDash
    &
    \constraint{[0,1,2] \mapsto [1,2,3]}
    &
    \constraint{[0,0] \mapsto [0]}
    &
    \constraint{[0,0,1] \mapsto [1,0,0]}
    \\[3pt]
    \hole{1} \vDash
    &
    \left\{
    \begin{array}{c}
      0 \mapsto 1 \\
      1 \mapsto 2 \\
      2 \mapsto 3
    \end{array}
    \right\}
    &
    \left\{
    \begin{array}{c}
      0 \mapsto 0 \\
      0 \mapsto \phantom{0} \\
    \end{array}
    \right\}
    &
    \left\{
    \begin{array}{c}
      0 \mapsto 1 \\
      0 \mapsto 0 \\
      1 \mapsto 0
    \end{array}
    \right\}
    \\
    & & = \bot & = \bot
  \end{array}
\]
  \caption{The inputs and outputs for example propagation through the hole
  filling \fbox{$\hole{0}~\mapsto~\fun{map}~\hole{1}$}, for example constraints
  taken from the functions \fun{inc}, \fun{compress} and \fun{reverse}. The
  latter two cannot be implemented using \fun{map}, which is reflected in the
  contradictory constraints: for \fun{compress} there is a length mismatch, and
  for \fun{reverse} the same input is mapped to different outputs.}
\label{fig:map}
\end{figure}

See Figure~\ref{fig:map} for examples of example propagation through \fun{map},
for various example constraints based on common functions on lists. The function
\fun{inc}, which increments each number in a list by one, can be implemented
using \fun{map}. As such, example propagation succeeds, resulting in a
constraint which represents incrementing a number by one. The function
\fun{compress}, which removes consecutive duplicates from a list, cannot be
implemented using \fun{map}, since the input and output lists can have
different lengths. As such, example propagation fails, as seen in
Figure~\ref{fig:map}. The function \fun{reverse}, which reverses a list, has
input and output lists of the same length. It can, however, not be implemented
using \fun{map}, as \fun{map} cannot take the positions of elements in a list
into account. This is reflected in the example in Figure~\ref{fig:map}, where
the resulting constraint is inconsistent, mapping $0$ to two different values.

For each combinator in \lam{}, a deduction rule is defined that captures the
various properties relevant for example propagation. This allows \lam{} to
efficiently synthesize complex functions in terms of these combinators. For
example, \lam{} is able to synthesize a function computing the Cartesian
product in terms of \fun{foldr}, believed to be the first functional
pearl~\cite{danvy_2007}.

\lam{} shows that synthesis using example propagation is feasible, but it is
not general purpose. Many synthesis problems require other recursion schemes or
are defined over different types. Example propagation can be added for other
functions in a similar fashion by adding new deduction rules, but this
is very laborious work.


\subsection{Example Propagation in \myth{}}

Osera and Zdancewic~\cite{osera_2015,osera_2016} take a more general approach
in their synthesizer \myth{}, compared to \lam{}, by focusing on structural
recursion, rather than a specific recursion scheme such as \fun{foldr}. To do so,
they describe how example constraints can be propagated through constructors
and pattern matches. Note that their language does not contain primitive
integers. Rather, literals $0$, $1$, $2$, etc.~are syntactic sugar for
Peano-style natural numbers.
\[
  \key{data}~ \typ{Nat} = \ctr{Zero}~ |~ \ctr{Succ}~ \typ{Nat}
\]

\subsubsection{Constructors}

To propagate a constraint through a constructor, we have to check that all
possible outputs agree with this constructor. For example, the constraint
$\constraint{\ctr{Zero}}$ can be refined by the constructor \ctr{Zero}. No
constraints need to be propagated, since \ctr{Zero} has no arguments. In the
next example, there are multiple possible outputs, depending on the value of
the variable $x$.
\[
  \hole{0} \vDash
  \left\{
  \begin{array}{c|l}
    x \\
    \dots & \ctr{Succ}~ \ctr{Zero} \\
    \dots & \ctr{Succ}~ (\ctr{Succ}~ \ctr{Zero})
  \end{array}
  \right\}
  \quad
  \xrightarrow{\ctr{Succ}}
  \quad
  \ctr{Succ}~(\hole{1} \vDash
  \left\{
  \begin{array}{c|l}
    x \\
    \dots & \ctr{Zero} \\
    \dots & \ctr{Succ}~ \ctr{Zero}
  \end{array}
  \right\}
  )
\]
Since every possible output is a successor, the constraint can be propagated
through \ctr{Succ} by removing one \ctr{Succ} constructor from each output,
i.e.~decreasing each output by one. The resulting constraint on $\hole{1}$
cannot be refined by a constructor, since the outputs do not all agree.


\subsubsection{Pattern Matching}

The elimination of constructors (i.e.~pattern matching) is a bit more
complicated. \myth{} describes example propagation through non-nested pattern
matches, as long as the scrutinee has no holes. Consider the following example,
wherein the sketch $\fun{double}=\lambda n .~ \hole{0}$ is refined by
propagating the constraint on $\hole{0}$ through a pattern match on the local
variable $n$.
\[
  \hole{0} \vDash
  \left\{
  \begin{array}{c|c}
    n \\
    0 & 0 \\
    1 & 2 \\
    2 & 4
  \end{array}
  \right\}
  \quad
  \xrightarrow{\textbf{pattern match}}
  \quad
  \begin{array}{l}
    \key{case}~ n~ \key{of} \\
    \quad
    \begin{array}{ll}
      \ctr{Zero}    &\rightarrow \hole{1} \vDash \constraint{0} \\[0.5mm]
      \ctr{Succ}~ m &\rightarrow \hole{2} \vDash
      \left\{
      \begin{array}{c|c}
        m \\
        0 & 2 \\
        1 & 4
      \end{array}
      \right\}
    \end{array}
  \end{array}
\]
Pattern matching on $n$ creates two branches, one for each constructor of
\typ{Nat}, with holes on the right-hand side. The constraint on $\hole{0}$ is
propagated to each branch by splitting up the constraint based on the value of
$n$. For brevity, we leave $n$ out of the new constraints. The newly introduced
variable $m$ is exactly one less than $n$, i.e.~one \ctr{Succ} constructor is
stripped away.


\subsubsection{Structural Recursion}

With support for structural recursion, \myth{} is able to perform
general-purpose, propagation-based synthesis. To illustrate this, we show how
\myth{} synthesizes the function \fun{double}, starting from the previous
sketches. Hole $\hole{1}$ is easily refined with \ctr{Zero}. Hole $\hole{2}$
can be refined with \ctr{Succ} twice, since every output is at least 2:
\[
  \hole{2} \vDash
  \left\{
  \begin{array}{c|c}
    m \\
    0 & 2 \\
    1 & 4
  \end{array}
  \right\}
  \quad
  \xrightarrow{\ctr{Succ}}
  \dots
  \xrightarrow{\ctr{Succ}}
  \quad
  \ctr{Succ}~ (\ctr{Succ}~ (\hole{3} \vDash
    \left\{
    \begin{array}{c|c}
      m \\
      0 & 0 \\
      1 & 2
    \end{array}
    \right\}
  ))
\]
At this point, to tie the knot, \myth{} should introduce the recursive call
$\fun{double}~ m$. Note, however, that \fun{double} is not yet implemented, so
we cannot directly test the correctness of this guess. We can, however, use the
original constraint (on $\hole{0}$) as a partial implementation of
\fun{double}. The example constraint on $\hole{3}$ is a subset of this original
constraint, with $m$ substituted for $n$. This implies that $\fun{double}~ m$
is a valid refinement. This property of example constraints, i.e.~that the
specification for recursive calls is a subset of the original constraint, is
known as \textit{trace completeness}~\cite{osera_2016}, and is a prerequisite
for synthesizing recursive functions in \myth{}.



\subsection{Example Propagation in \smyth{}}

In their synthesizer \smyth{}, Lubin et al.~\cite{lubin_2020} extend \myth{}
with sketching, i.e.~program synthesis starting from a sketch, a program
containing holes. A global example constraint is propagated through the sketch
using example propagation, after which \myth{}-style synthesis takes over using
the local example constraints.

Take for example the constraint $\constraint{[0,1,2] \mapsto [0,2,4]}$, which
represents doubling each number in a list. The programmer may provide the
sketch $\fun{map}~ \hole{}$ as a starting point for the synthesis procedure.
In order to perform \myth{}-style synthesis, the constraint has to be
propagated through \fun{map}, but unlike \lam{}, \smyth{} does not provide a
handcrafted rule for \fun{map}. Instead, \smyth{} determines how examples are
propagated through functions based on their implementation.

The crucial idea is that the sketch is first evaluated, essentially inlining
all function calls\footnote{Note that function calls within the branches of a
stuck pattern match are not inlined.} until only simple language constructs
remain, each of which supports example propagation. Omar et
al.~\cite{omar_2019} describe how to evaluate an expression containing holes
using live evaluation. The sketch is applied to the provided input, after which
\fun{map} is inlined and evaluated as far as possible:
\[
  \fun{map}~ \hole{}~ [0,1,2]
  \hfill
  \leadsto
  \hfill
  [\hole{}~ 0,~ \hole{}~ 1,~ \hole{}~ 2]
\]
At this point, the constraint can be propagated through the resulting
expression. Lubin et al.~\cite{lubin_2020} extend \myth{}-style example
propagation to work for the primitives returned by live evaluation. The
constraint is propagated through the result of live evaluation.
\[
  \begin{array}{l}
    [\hole{}~ 0,~ \hole{}~ 1,~ \hole{}~ 2]
    \vDash
    \constraint{[0, 2, 4]}
    \quad
    \longrightarrow^{*}
    \quad
    \begin{array}{rl}
      [ &(\hole{} \vDash \constraint{0 \mapsto 0})~ 0 \\
      , &(\hole{} \vDash \constraint{1 \mapsto 2})~ 1 \\
      , &(\hole{} \vDash \constraint{2 \mapsto 4})~ 2~ ]
    \end{array}
  \end{array}
\]
The constraints propagated to the different occurrences of $\hole{}$ in the
evaluated expression can then be collected and combined to compute a constraint
for $\hole{}$ in the input sketch.
\[
  \fun{map}~ (\hole{} \vDash
  \left\{
  \begin{array}{l}
    0 \mapsto 0 \\
    1 \mapsto 2 \\
    2 \mapsto 4
  \end{array}
  \right\}
  )
\]
This kind of example propagation based on evaluation is called live
bidirectional evaluation. For a full description, see Lubin et
al.~\cite{lubin_2020}. \smyth{} uses live bidirectional evaluation to extend
\myth{} with sketching. Note, however, that \smyth{} does not use live
bidirectional evaluation to introduce refinements during synthesis.




\section{Program Synthesis Using Example Propagation}\label{sec:techniques}

We define our synthesis problem as finding an expression of type $\tau$ in the
environment $\Gamma$ that adheres to example constraint $\varphi$. Inspired by
Smith and Albarghouthi~\cite{smith_2019}, we give a high-level overview of the
synthesis procedure as a set of guarded rules that can be applied
non-deterministically, shown in Figure~\ref{fig:rules}. We keep track of a set
of candidate expressions $\mathcal{E}$, which is initialized by the rule
\textsc{init} and then expanded by the rule \textsc{expand} until the rule
\textsc{final} applies, returning a solution.

The rule \textsc{init} initializes $\mathcal{E}$ with a single hole $\hole{0}$,
constrained by the synthesis parameters. Each invocation of the rule
\textsc{expand} non-deterministically picks an expression $e$ from
$\mathcal{E}$ and a hole $\hole{i}$ in $e$ to fill, by generating a hole
filling \textit{hf} using the context and type of $\hole{i}$. If the resulting
expression $e'$ does not conflict with $\varphi$, it is considered a valid
candidate and added to $\mathcal{E}$. As an invariant, $\mathcal{E}$ only
contains expressions that do not conflict with $\varphi$. As such, a solution
to the synthesis problem is simply any expression that has no holes.

To implement a synthesizer according to these rules, we have to make the
non-deterministic choices explicit: we have to decide in which order
expressions are expanded ($e \in \mathcal{E}$); which holes are selected for
expansion ($\hole{} \in \textit{holes}(e)$); and how hole
fillings are generated based on the hole's type and environment ($\Gamma \vdash
\textit{hf} : \tau$). Additionally, we describe how expressions containing
holes are checked against the constraint $\varphi$.

\begin{figure}[t]
\centering
  \begin{minipage}{0.5\textwidth}
    \centering
    \[
      \inference
      { \Gamma \vdash \hole{0} : \tau \vDash \varphi
      }
      {\mathcal{E} \leftarrow \{ \hole{0} \}}[\textsc{init}]
    \]
    \\[-10pt]
    \[
      \inference
      {
        e \in \mathcal{E}
        &
        \hole{i} \in \textit{holes}(e)
        &
        \Gamma_{i} \vdash \textit{hf} : \tau_{i}
        \\
        e' = [\hole{i} \mapsto \textit{hf} \; ]  e
        \quad\quad
        e' \vDash \varphi
      }
      {
        \mathcal{E} \leftarrow \mathcal{E} \cup \{ e' \}
      }
      [\textsc{expand}]
    \]
    \\[-10pt]
    \[
      \inference
      {
        e \in \mathcal{E}
        &
        \textit{holes}(e) = \emptyset
      }
      {e~\text{is a solution}}
      [\textsc{final}]
    \]
    \caption{Program synthesis using example propagation as a set of guarded
    rules that can be applied non-deterministically.}
    \label{fig:rules}
  \end{minipage}\hfill
  \begin{minipage}{0.4\textwidth}
    \centering
    \[
      \boxed{
        \begin{array}{l}
          \hole{0} \mapsto \fun{map}~ (\lambda x .~ \hole{1})~ \hole{2} \\
          \hole{1} \mapsto \ctr{Succ}~ \hole{3} \\
          \hole{2} \mapsto xs \\
          \hole{3} \mapsto x
        \end{array}
      }
    \]
    \caption{A set of hole fillings synthesizing an expression that increments
    each value in a list, starting from the sketch $\lambda xs .~ \hole{0}$.}
    \label{fig:fillings}
  \end{minipage}
\end{figure}


\subsection{Expression Order}

To decide in which order candidate expressions are selected for expansion, we
define an order on expressions by assigning a weight to each expression. We
keep track of all expressions in a priority queue and expand expressions in
increasing order of their weight. The weight of an expression is computed as
follows: we assign a weight of 1 to each application, as well as to each
pattern match and each call to a recursion scheme. Additionally, the weight of
scrutinees is doubled, to disincentivize pattern matching on large scrutinees.


\subsection{Hole Order}

Choosing in which order holes are filled during synthesis is a bit more
involved. Consider, for example, the hole fillings in
Figure~\ref{fig:fillings}, synthesizing the expression $\lambda xs .~
\fun{map}~ (\lambda x .~ \ctr{Succ}~ x)~ xs$ starting from $\lambda xs .~
\hole{0}$. There are three different synthesis paths that lead to this result,
depending on which holes are filled first. More specifically, $\hole{2}$ can be
filled independently of $\hole{1}$ and $\hole{3}$, so it could be filled
before, between, or after them. To avoid generating the same expression three
times, we should fix the order in which holes are filled, so that there is a
unique path to every possible expression.


Because our techniques rely heavily on evaluation, we let evaluation guide the
hole order. After filling hole $\hole{0}$, we live evaluate.
\[
  \fun{map}~ (\lambda x .~ \hole{1})~ \hole{2}
  \leadsto
  \key{case}~ \hole{2}~ \key{of}~ \dots
\]
At this point, evaluation cannot continue, because we do not know
which pattern $\hole{2}$ will be matched on. We say that $\hole{2}$ blocks the
evaluation. By filling $\hole{2}$, the pattern match may resolve and generate
new example constraints for $\hole{1}$. Conversely, filling $\hole{1}$ does not
introduce any new constraints. Hence, we always fill blocking holes first.
Blocking holes are easily computed by live evaluating the expression against
the example constraints.





\subsection{Generating Hole Fillings}\label{sec:betaeta}

Hole fillings depend on the local context and the type of a hole and may
consist of constructors, pattern matches, variables and function calls. To
avoid synthesizing multiple equivalent expressions, we will only generate
expressions in \betaeta{} form. An expression is in \betaeta{} form exactly if
no $\eta$-expansions or $\beta$-reductions are possible. During synthesis, we
guarantee \betaeta{} form by greedily $\eta$-expanding newly introduced holes
and always fully applying functions, variables and constructors. Consider, for
example, the function \fun{map}. To use \fun{map} as a refinement, it is
applied to two holes, the first of which is $\eta$-expanded:
\[
  \fun{map}~ (\lambda x.~ \hole{0})~ \hole{1}
\]
Pattern matches can be handled in the same way by interpreting them as
eliminator functions, which are equivalent in expressiveness.

Furthermore, we add some syntactic restrictions to the generated expressions:
we only allow the recursive argument of recursion schemes such as \fun{foldr}
to be variables. This is similar to the restriction on structural recursion in
\myth{}~\cite{osera_2015,osera_2016} and \smyth{}~\cite{lubin_2020}.
Additionally, we disallow expressions that are not in normal form, somewhat
similar to equivalence reduction as described by Smith and
Albarghouthi~\cite{smith_2019}. Currently, our tool provides a handcrafted set
of expressions that are not in normal form, which are prohibited during
synthesis. Ideally, these sets of disallowed expressions would be taken from an
existing data set (such as
HLint\footnote{\url{https://github.com/ndmitchell/hlint}}), or approximated
using evaluation-based techniques such as
\textsc{QuickSpec}~\cite{claessen_2010}.

\subsection{Pruning Expressions}

For an expression $e \in \mathcal{E}$ and a hole $\hole{i} \in
\textit{holes}(e)$, we generate a set of possible hole fillings based on the
hole context $\Gamma_i$ and the hole type $\tau_i$. For each of these hole
fillings, we try to apply the \textsc{expand} rule. To do so, we must check
that the resulting expression $e'$ does not conflict with the example
constraint $\varphi$. We use \smyth{}-style example propagation to compute hole
constraints for $e'$. If example propagation fails, we do not add $e'$ to
$\mathcal{E}$, essentially pruning the search space.



\subsubsection{Diverging Constraints}\label{sec:diverge}

Unfortunately, example propagation is not feasible for all possible
expressions. Consider, for instance, the function \fun{sum}. If we try to
propagate a constraint through $\fun{sum}~ \hole{}$, we first use live
evaluation, resulting in the following partially evaluated result, with
$\hole{}$ in a scrutinized position:
\[
  \begin{array}{l}
    \key{case}~ \hole{}~ \key{of} \\
    \quad
    \begin{array}{ll}
      [\:] &\rightarrow 0 \\
      x:xs &\rightarrow \fun{plus}~ x~ (\fun{sum}~ xs)
    \end{array}
  \end{array}
\]
Unlike \myth{}, \smyth{} allows examples to be propagated through pattern
matches whose scrutinee may contain holes, considering each branch separately
under the assumption that the scrutinee evaluates to the corresponding pattern.
This introduces disjunctions in the example constraint. Propagating
$\constraint{\ctr{Zero}}$ through the previous expression results in a
constraint that cannot be finitely captured in our constraint language:
\[
  (\hole{} \vDash \constraint{[\:]})
  \lor
  (\hole{} \vDash \constraint{[\ctr{Zero}]})
  \lor
  (\hole{} \vDash \constraint{[\ctr{Zero},\ctr{Zero}]})
  \lor
  \dots
\]
Without extending the constraint language it is impossible to compute such a
constraint. Instead, we try to recognize that example propagation diverges, by
setting a maximum to the amount of recursive calls allowed during example
propagation. If the maximum recursion depth is reached, we cancel example
propagation.

Since example propagation through $\fun{sum}~ \hole{}$ always diverges, we
could decide to disallow it as a hole filling. This is, however, too
restrictive, as example propagation becomes feasible again when the length of
the argument to \fun{sum} is no longer unrestricted.
Take, for example, the following constraint, representing counting the number
of \ctr{True}s in a list, and a possible series of hole fillings:
\[
  \hole{0} \vDash
  \left\{
  \begin{array}{c|c}
    xs \\[0mm]
    [\ctr{False}] & 0 \\[0mm]
    [\ctr{False},\ctr{True}] & 1 \\[0mm]
    [\ctr{True},\ctr{True}] & 2
  \end{array}
  \right\}
  \quad\quad\quad\quad
  \boxed{
    \begin{array}{l}
      \hole{0} \mapsto \fun{sum}~ \hole{1} \\
      \hole{1} \mapsto \fun{map}~ (\lambda x .~ \hole{2})~ \hole{3} \\
      \hole{3} \mapsto xs \\
    \end{array}
  }
\]
Trying to propagate through $\boxed{\hole{0} \mapsto \fun{sum}~ \hole{1}}$
diverges, since $\hole{1}$ could be a list of any length. At this point, we
could decide to disregard this hole filling, but this would incorrectly prune
away a valid solution. Instead, we allow synthesis to continue guessing hole
fillings, until we get back on the right track: after guessing
$\boxed{\hole{1}\mapsto\fun{map}~(\lambda x.~\hole{2})~\hole{3}}$ and
$\boxed{\hole{3}\mapsto xs}$, the length of the argument to \fun{sum} becomes
restricted and example propagation no longer diverges:
\[
  \fun{sum}~ (\fun{map}~ (\lambda x .~ \hole{2} \vDash
  \left\{
  \begin{array}{c|c}
    x \\
    \ctr{False} & 0 \\
    \ctr{True}  & 1
  \end{array}
  \right\}
  )~ xs)
\]
At this point, synthesis easily finishes by pattern matching on $x$. Note that,
unlike \lam{}, \myth{} and \smyth{}, \scrybe{} is able to interleave
refinements and guesses.

\subsubsection{Exponential Constraints}

Even if example propagation does not diverge, it still might take too long to
compute or generate a disproportionally large constraint, slowing down the
synthesis procedure. Lubin et al.~\cite{lubin_2020} compute the falsifiability
of an example constraint by first transforming it to disjunctive normal form
(DNF), which may lead to exponential growth of the constraint size. For
example, consider the function \fun{or}, defined as follows:
\[
  \begin{array}{l}
    \fun{or} = \lambda a~ b .~ \key{case}~ a~ \key{of} \\
    \quad
    \begin{array}{ll}
      \ctr{False} &\rightarrow b \\
      \ctr{True}  &\rightarrow \ctr{True}
    \end{array}
  \end{array}
\]
Propagating the example constraint $\constraint{\ctr{True}}$ through the
expression $\fun{or}~ \hole{0}~ \hole{1}$ puts the hole $\hole{0}$ in a
scrutinized position, resulting in the following constraint:
\[
  (\hole{0} \vDash \constraint{\ctr{False}}
  \land \hole{1} \vDash \constraint{\ctr{True}})
  \lor \hole{0} \vDash \constraint{\ctr{True}}
\]
This constraint has size three (the number of hole occurrences). We can extend this
example by mapping it over a list of length $n$ as follows:
\[
  \fun{map}~ (\lambda x .~ \fun{or}~ \hole{0}~ \hole{1})~ [0,1,2, \dots]
  \vDash
  \constraint{[\ctr{True}, \ctr{True}, \ctr{True}, \dots]}
\]
Propagation generates a conjunction of $n$ constraints that are all exactly the
same apart from their local context, which differs in the value of $x$. This
constraint, unsurprisingly, has size $3n$. Computing the disjunctive normal
form of this constraint, however, results in a constraint of size of $2^n
\times \frac{3}{2}n$, which is exponential.

In some cases, generating such a large constraint may cause example propagation
to reach the maximum recursion depth. In other cases, example propagation
succeeds, but returns such a large constraint that subsequent refinements will
take too long to compute. In both cases, we treat it the same as diverging
example propagation.







\section{Evaluation}\label{sec:benchmark}


To evaluate \scrybe{}, we combine the benchmarks of
\myth{}~\cite{osera_2015} and \lam{}~\cite{feser_2015}. This evaluation is not
intended to compare our technique directly with previous techniques in terms of
efficiency, but rather to show the wide range of synthesis problems that
\scrybe{} can handle. Additionally, we get some insight in the effectiveness of
example propagation as a pruning technique.

For ease of readability, the benchmark suite is split up into a set of
functions operating on lists (Table~\ref{tab:listbench},
Appendix~\ref{sec:appendix}) and a set of functions operating on binary trees
(Table~\ref{tab:treebench}, Appendix~\ref{sec:appendix}). We have excluded
functions operating on just booleans or natural numbers, as these are all
trivial and synthesize in a few milliseconds. For consistency, and to avoid
naming conflicts, the names of some of the benchmarks are changed to reflect
the corresponding functions in the Haskell prelude. To avoid confusion, each
benchmark function comes with a short description.

Each row describes a single synthesis problem in terms of a function that
needs to be synthesized. The first two columns give the name and a short
description of this function. The third and fourth columns show, in
milliseconds, the average time our synthesizer takes to correctly synthesize
the function with example propagation (\textit{EP}) and without example
propagation (\textit{NoEP}), respectively. Some functions may fail to
synthesize ($\bot$) within 5 seconds and some cannot straightforwardly be
represented in our language (-). The last three columns show, for \myth{},
\smyth{}, and \lam{}, respectively, whether the function synthesizes (\yes),
fails to synthesize (\no), or is not included in their benchmark (-).

The benchmarks list\_head, list\_tail, list\_init and list\_last are all
partial functions (marked $\dagger$). We do not support partial functions, and
therefore these functions are replaced by their total equivalents, by wrapping
their return type in \typ{Maybe}. For example, list\_last is defined as
follows, where the outlined hole filling is the result returned by \scrybe{}
(input-output constraints are omitted for brevity):
\[
  \begin{array}{l}
    \{-\#~ \text{USE}~ \fun{foldr} ~\#-\} \\[2pt]
    \fun{list\_last} :: \typ{List}~ a \rightarrow \typ{Maybe}~ a \\
    \fun{list\_last}~ \textit{xs} = \hole{} \\[3pt]
    \boxed{\hole{} \mapsto \fun{foldr}~ (\lambda x~ r.~ \key{case}~ r~ \key{of}
    \;
    \begin{array}{ll}
      \ctr{Nothing} &\rightarrow \ctr{Just}~ x \\[-1pt]
      \ctr{Just}~ y &\rightarrow r
    \end{array}
    )~ \ctr{Nothing}~ \textit{xs}}
  \end{array}
\]
The benchmarks list\_drop, list\_index, list\_take and tree\_level (marked
$\ddagger$) all recurse over two datatypes at the same time. As such, they
cannot be implemented using \fun{foldr} as it is used in
Section~\ref{sec:betaeta}. Instead, we provide a specialized version of foldr
that takes an extra argument:
\[
  \begin{array}{l}
    \{-\#~ \text{USE}~ \fun{foldr} :: (a \rightarrow (c \rightarrow b)
    \rightarrow (c \rightarrow b)) \rightarrow (c \rightarrow b) \rightarrow
    \typ{List}~ a \rightarrow c \rightarrow b ~\#-\} \\[2pt]
    \fun{list\_take} :: \typ{Nat} \rightarrow \typ{List}~ a \rightarrow \typ{List}~ a \\
    \fun{list\_take}~ n~ \textit{xs} = \hole{} \\[3pt]
    \boxed{\hole{} \mapsto \fun{foldr}~ (\lambda x~ r~ m.~ \key{case}~ m~
    \key{of}
    \;
    \begin{array}{ll}
      \ctr{Zero}    &\rightarrow [\:] \\[-1pt]
      \ctr{Succ}~ o &\rightarrow (x:r~o)
    \end{array}
    )~ (\lambda \_.~ [\:])~ \textit{xs}~ n}
  \end{array}
\]
A few functions (marked *) could not straightforwardly be translated to our
approach:
\begin{itemize}
  \item Function list\_delete\_mins requires a total function in scope that
    returns the minimum number in a list. This is not possible for natural
    numbers, as there is no obvious number to return for empty lists.
  \item Function list\_swap uses nested pattern matching on the input list,
    which is not possible to mimic using a fold.
  \item Function list\_reverse combines a set of benchmarks from \myth{} that
    synthesize \fun{reverse} using different techniques, which are not easily
    translated to our language.
\end{itemize}

\subsection{Results}

\scrybe{} is able to synthesize most of the combined benchmarks of \myth{} and
\lam{}, with a median runtime of 15.95 milliseconds. Furthermore, synthesis
with example propagation is on average 5.22 times as fast as without example
propagation, disregarding the benchmarks where synthesis without example
propagation failed. \lam{} noticed a similar improvement (6 times as fast)
for example propagation based on automated deduction, which indicates that
example propagation using live-bidirectional evaluation is similar in strength,
while being more general.

Some functions benefit especially from example propagation, in particular
problems that are composed of multiple synthesis problems. Take, for example,
tree\_snoc, which effectively synthesizes \fun{mapTree} and \fun{snoc} from
\fun{foldTree} and \fun{foldr} respectively. Without example propagation, it is
not tractable to automatically decompose this synthesis problem into these two
parts.
\[
  \begin{array}{l}
    \{-\#~ \text{USE}~ \fun{foldTree}, \fun{foldr} ~\#-\} \\[2pt]
    \fun{tree\_snoc} :: a \rightarrow \typ{Tree}~ (\typ{List}~ a) \rightarrow
      \typ{Tree}~ (\typ{List}~ a) \\
    \fun{tree\_snoc}~ x~ t = \hole{} \\[3pt]
    \boxed{\hole{} \mapsto \fun{foldTree}~
      (\lambda l~ \textit{xs}~ r.~ \ctr{Node}~ l~
      (\fun{foldr}~ (\lambda x~ q.~ x:q)~ [x]~ \textit{xs})~ r)~ \ctr{Leaf}~ t}
  \end{array}
\]
On the other hand, for some functions, such as tree\_search, synthesis is
noticeably faster without example propagation, showing that the overhead of
example propagation sometimes outweighs the benefits. This indicates that it
might be helpful to use some heuristics to decide when example propagation is
beneficial.
A few functions that fail to synthesize, such as list\_compress, do synthesize
when a simple sketch is provided:
\[
  \begin{array}{ll}
    \begin{array}{l}
      \{-\#~ \text{USE}~ \fun{foldr}, (\equiv) ~\#-\} \\[2pt]
      \fun{compress} :: \typ{List}~ \typ{Nat} \rightarrow \typ{List}~ \typ{Nat} \\
      \fun{compress}~ xs = \fun{foldr}~ (\lambda x~r.~\hole{0})~\hole{1}
    \end{array}
    &
    \boxed{
      \begin{array}{l}
        \hole{0} \mapsto [\:]
        \\
        \hole{1} \mapsto x : \key{case}~ r~ \key{of} \\
        \quad
        \begin{array}{ll}
          [\:] &\rightarrow r \\[-1pt]
          y:ys &\rightarrow
          \key{if}~ x \equiv y~ \key{then}~ ys~ \key{else}~ r
        \end{array}
      \end{array}
    }
  \end{array}
\]
Since our evaluation was not aimed at sketching, we still consider
list\_compress to fail ($\bot$).


\section{Conclusion}




We presented an approach to program synthesis using example propagation that
specializes in compositionality, by allowing arbitrary functions to be used as
refinement steps. One of the key ideas is holding on to constraint information
as long as possible, rather than resorting to brute-force, enumerative search.
Our experiments show that we are able to synthesize a wide range of synthesis
problems from different synthesis domains.

There are many avenues for future research. One direction we wish to explore is
to replace the currently ad hoc constraint solver with a more general purpose
SMT solver. Our hope is that this paves the way for the addition of primitive
data types such as integers and floating point numbers.










\subsubsection*{Acknowledgements} We would like to thank Alex Gerdes, Koen
Claessen, and the anonymous reviewers of HATRA 2022 for their supportive
comments and constructive feedback.

\bibliographystyle{splncs04}
\bibliography{citations}

\begin{thebibliography}{10}
\providecommand{\url}[1]{\texttt{#1}}
\providecommand{\urlprefix}{URL }
\providecommand{\doi}[1]{https://doi.org/#1}

\bibitem{claessen_2010}
Claessen, K., Smallbone, N., Hughes, J.: Quickspec: Guessing formal
  specifications using testing. In: Fraser, G., Gargantini, A. (eds.) Tests and
  Proofs. pp. 6--21. Springer Berlin Heidelberg (2010)

\bibitem{danvy_2007}
Danvy, O., Spivey, M.: On {B}arron and {S}trachey's cartesian product function.
  In: Proceedings of the 12th ACM SIGPLAN International Conference on
  Functional Programming. p. 41–46. ICFP '07, Association for Computing
  Machinery, New York, NY, USA (2007),
  \url{https://doi.org/10.1145/1291151.1291161}

\bibitem{feser_2015}
Feser, J.K., Chaudhuri, S., Dillig, I.: Synthesizing data structure
  transformations from input-output examples. ACM SIGPLAN Notices
  \textbf{50}(6),  229--239 (Aug 2015),
  \url{https://dl.acm.org/doi/10.1145/2813885.2737977}

\bibitem{guo_2019}
Guo, Z., James, M., Justo, D., Zhou, J., Wang, Z., Jhala, R., Polikarpova, N.:
  Program synthesis by type-guided abstraction refinement. Proc. ACM Program.
  Lang.  \textbf{4}({POPL}) (Dec 2019), \url{https://doi.org/10.1145/3371080}

\bibitem{katayama_2008}
Katayama, S.: Efficient exhaustive generation of functional programs using
  monte-carlo search with iterative deepening. In: Ho, T.B., Zhou, Z.H. (eds.)
  PRICAI 2008: Trends in Artificial Intelligence. pp. 199--210. Springer Berlin
  Heidelberg (2008)

\bibitem{koppel_2022}
Koppel, J., Guo, Z., de~Vries, E., Solar-Lezama, A., Polikarpova, N.: Searching
  entangled program spaces. Proc. ACM Program. Lang.  \textbf{6}({ICFP}) (aug
  2022), \url{https://doi.org/10.1145/3547622}

\bibitem{lubin_2020}
Lubin, J., Collins, N., Omar, C., Chugh, R.: Program sketching with live
  bidirectional evaluation. Proc. ACM Program. Lang.  \textbf{4}({ICFP}) (Aug
  2020), \url{https://doi.org/10.1145/3408991}

\bibitem{omar_2019}
Omar, C., Voysey, I., Chugh, R., Hammer, M.A.: Live functional programming with
  typed holes. Proc. ACM Program. Lang.  \textbf{3}({POPL}) (Jan 2019),
  \url{https://doi.org/10.1145/3290327}

\bibitem{osera_2016}
Osera, P.M.: Programming assistance for type-directed programming (extended
  abstract). In: Proceedings of the 1st {International} {Workshop} on
  {Type}-{Driven} {Development} - {TyDe} 2016. pp. 56--57. ACM Press, Nara,
  Japan (2016), \url{http://dl.acm.org/citation.cfm?doid=2976022.2976027}

\bibitem{osera_2015}
Osera, P.M., Zdancewic, S.: Type-and-example-directed program synthesis. ACM
  SIGPLAN Notices  \textbf{50}(6),  619--630 (Aug 2015),
  \url{https://dl.acm.org/doi/10.1145/2813885.2738007}

\bibitem{peleg_2020}
Peleg, H., Gabay, R., Itzhaky, S., Yahav, E.: Programming with a
  read-eval-synth loop. Proc. ACM Program. Lang.  \textbf{4}({OOPSLA}) (nov
  2020), \url{https://doi.org/10.1145/3428227}

\bibitem{smith_2019}
Smith, C., Albarghouthi, A.: Program synthesis with equivalence reduction. In:
  Enea, C., Piskac, R. (eds.) Verification, Model Checking, and Abstract
  Interpretation. pp. 24--47. Springer International Publishing, Cham (2019)

\bibitem{solar-lezama_2008}
Solar-Lezama, A.: Program Synthesis by Sketching. Ph.D. thesis, Berkeley (2008)

\bibitem{solar-lezama_2009}
Solar-Lezama, A.: The sketching approach to program synthesis. In: Hu, Z. (ed.)
  Programming Languages and Systems. pp. 4--13. Springer Berlin Heidelberg
  (2009)

\end{thebibliography}

\appendix



\begin{table*}[htbp]
\section{Appendix}\label{sec:appendix}

\hspace*{-2cm}
\setlength{\tabcolsep}{2pt}
\begin{tabular}{ l|l|r|r|c c c }

  {\it Function} &
  {\it Description} &
  {\it EP (ms) } &
  {\it NoEP (ms) } &
  \myth{} &
  \smyth{} &
  \lam{} \\[2mm]

  list\_add
  & Increment each value in a list by $n$
  & 4.96
  & 11.60
  & - & - & \yes \\

  \rowcolor{highlight}
  list\_append
  & Append two lists
  & 4.90
  & 18.30
  & \yes & \yes & \yes \\

  list\_cartesian
  & The cartesian product
  & 449.00
  & $\bot$
  & - & - & \yes \\

  \rowcolor{highlight}
  list\_compress
  & Remove consecutive duplicates from a list
  & $\bot$ 
  & $\bot$
  & \yes & \no & - \\

  list\_flatten
  & Flatten a list of lists
  & 3.53
  & 3.00
  & \yes & \yes & \yes \\

  \rowcolor{highlight}
  list\_copy\_first
  & Replace each element in a list with the first
  & 40.10
  & 82.90
  & - & - & \yes \\

  list\_copy\_last
  & Replace each element in a list with the last
  & 38.20
  & 106.00
  & - & - & \yes \\

  \rowcolor{highlight}
  list\_delete\_max
  & Remove the largest numbers from a list
  & 38.60
  & 127.00
  & - & - & \yes \\

  list\_delete\_mins*
  & Remove the smallest numbers from a list of lists
  & -
  & -
  & - & - & \yes \\

  \rowcolor{highlight}
  list\_drop\indexed
  & All but the first $n$ elements of a list
  & 192.00
  & 473.00
  & \yes & \yes & - \\

  list\_even\_parity
  & Whether a list has an odd number of $True$s
  & 15.30
  & 99.90
  & \yes & \no & - \\

  \rowcolor{highlight}
  list\_evens
  & Remove the odd numbers from a list
  & 2.00
  & 4.60
  & - & - & \yes \\

  list\_filter
  & The elements in a list that satisfy $p$
  & 8.88
  & 46.30
  & \yes & \yes & - \\

  \rowcolor{highlight}
  list\_fold
  & A catamorphism over a list
  & 8.40
  & 5.55
  & \yes & \yes & - \\

  list\_head\total
  & The first element of a list
  & 0.65
  & 0.70
  & \yes & \yes & - \\

  \rowcolor{highlight}
  list\_inc
  & Increment each value in a list by one
  & 3.30
  & 221.00 
  & \yes & \yes & - \\

  list\_incs
  & Increment each value in a list of lists by one
  & 9.16
  & 23.40
  & - & - & \yes \\

  \rowcolor{highlight}
  list\_index\indexed
  & Index a list starting at 0
  & 51.80
  & 167.00
  & \yes & \yes & - \\

  list\_init\total
  & All but the last element of a list
  & 869.00
  & $\bot$
  & - & - & \yes \\

  \rowcolor{highlight}
  list\_last\total
  & The last element of a list
  & 167.00
  & 123.00
  & \yes & \yes & \yes \\

  list\_length
  & The number of elements in a list
  & 0.62
  & 2.86
  & \yes & \yes & \yes \\

  \rowcolor{highlight}
  list\_map
  & Map a function over a list
  & 1.38
  & 2.03
  & \yes & \yes & - \\

  list\_maximum
  & The largest number in a list
  & 26.20
  & 303.00
  & - & - & \yes \\

  \rowcolor{highlight}
  list\_member
  & Whether a number occurs in a list
  & 873.00
  & 4090.00
  & - & - & \yes \\

  list\_nub
  & Remove duplicates from a list
  & $\bot$
  & $\bot$
  & - & - & \yes \\

  \rowcolor{highlight}
  list\_swap*
  & Swap the elements in a list pairwise
  & -
  & -
  & \yes & \yes & - \\

  list\_reverse*
  & Reverse a list
  & 1.67
  & 2.57
  & \yes & \yes & - \\





  \rowcolor{highlight}
  list\_shiftl
  & Shift all elements in a list to the left
  & 69.00
  & 366.00
  & - & - & \yes \\

  list\_shiftr
  & Shift all elements in a list to the right
  & 89.20
  & 708.00
  & - & - & \yes \\

  \rowcolor{highlight}
  list\_snoc
  & Add an element to the end of a list
  & 69.00
  & 366.00
  & \yes & \yes & \yes \\

  list\_set\_insert
  & Insert an element in a set
  & $\bot$
  & $\bot$
  & \yes & \yes & - \\

  \rowcolor{highlight}
  list\_dupli
  & Duplicate each element in a list
  & 2.44
  & 3.33
  & \yes & \yes & \yes \\

  list\_sum
  & The sum of all numbers in a list
  & 3.59
  & 19.20
  & \yes & \yes & \yes \\

  \rowcolor{highlight}
  list\_sums
  & The sum of each nested list in a list of lists
  & 607.00
  & $\bot$
  & - & - & \yes \\

  list\_tail\total
  & All but the first element of a list
  & 0.85
  & 1.33
  & \yes & \yes & - \\

  \rowcolor{highlight}
  list\_take\indexed
  & The first $n$ elements of a list
  & 182.00
  & 3690.00
  & \yes & \yes & - \\

  list\_to\_set
  & Sort a list, removing duplicates
  & 458.00
  & $\bot$
  & \yes & \yes & - \\

\end{tabular}
\\
\caption{Benchmark for functions acting on lists. Each row describes a single
benchmark task and the time it takes for each function to synthesize with
example propagation (\textit{EP}) and without (\textit{NoEP}) respectively.
Some tasks cannot be synthesized within 5 seconds ($\bot$) and others are
omitted, since they cannot straightforwardly be translated to our language
(-).}
\label{tab:listbench}
\end{table*}
\vspace*{-6cm}

\begin{table*}[ht]
\hspace*{-2.2cm}
\setlength{\tabcolsep}{2pt}
\begin{tabular}{ l|l|r|r|c c c }

  {\it Function} &
  {\it Description} &
  {\it EP (ms) } &
  {\it NoEP (ms) } &
  \myth{} &
  \smyth{} &
  \lam{} \\[2mm]



  tree\_cons
  & Add an element to the front of each node in a tree of lists
  & 6.92
  & $\bot$
  & - & - & \yes \\

  \rowcolor{highlight}
  tree\_flatten
  & Flatten a tree of lists into a list
  & 20.80
  & 25.20
  & - & - & \yes \\

  tree\_height
  & The height of a tree
  & 7.71
  & 27.00
  & - & - & \yes \\

  \rowcolor{highlight}
  tree\_inc
  & Increment each element in a tree by one
  & 6.14
  & $\bot$
  & - & - & \yes \\

  tree\_inorder
  & Inorder traversal of a tree
  & 11.50
  & 9.17
  & \yes & \yes & \yes \\

  \rowcolor{highlight}
  tree\_insert
  & Insert an element in a binary tree
  & $\bot$
  & $\bot$
  & \yes & \no & - \\

  tree\_leaves
  & The number of leaves in a tree
  & 24.50
  & 40.00
  & \yes & \yes & \yes \\

  \rowcolor{highlight}
  tree\_maximum
  & The largest number in a tree
  & 31.90
  & 157.00
  & - & - & \yes \\

  tree\_map
  & Map a function over a tree
  & 2.61
  & 6.84
  & \yes & \yes & - \\

  \rowcolor{highlight}
  tree\_member
  & Whether a number occurs in a tree
  & 597.00
  & $\bot$
  & - & - & \yes \\

  tree\_level\indexed
  & The number of nodes at depth $n$
  & $\bot$
  & $\bot$
  & \yes & \no & - \\

  \rowcolor{highlight}
  tree\_postorder
  & Postorder traversal of a tree
  & 19.60
  & 24.40
  & \yes & \no & - \\

  tree\_preorder
  & Preorder traversal of a tree
  & 7.49
  & 15.40
  & \yes & \yes & - \\


  \rowcolor{highlight}
  tree\_search
  & Whether a number occurs in a tree of lists
  & 964.00
  & 307.00
  & - & - & \yes \\

  tree\_select
  & All nodes in a tree that satisfy $p$
  & 773.00
  & 3170.00
  & - & - & \yes \\

  \rowcolor{highlight}
  tree\_size
  & The number of nodes in a tree
  & 16.60
  & 39.50
  & \yes & \yes & \yes \\

  tree\_snoc
  & Add an element to the end of each node in a tree of lists
  & 81.60
  & $\bot$
  & - & - & \yes \\

  \rowcolor{highlight}
  tree\_sum
  & The sum of all nodes in a tree
  & 16.70
  & 104.00
  & - & - & \yes \\

  tree\_sum\_lists
  & The sum of each list in a tree of lists
  & 7.13
  & $\bot$
  & - & - & \yes \\

  \rowcolor{highlight}
  tree\_sum\_trees
  & The sum of each tree in a list of trees
  & 28.40
  & $\bot$
  & - & - & \yes \\

\end{tabular}
\\
\caption{Benchmark for functions acting on binary trees. Each row describes a
single benchmark task and the time it takes for each function to synthesize
with example propagation (\textit{EP}) and without (\textit{NoEP})
respectively. Some tasks cannot be synthesized within 5 seconds ($\bot$) and
others are omitted, since they cannot straightforwardly be translated to our
language (-).}
\label{tab:treebench}
\end{table*}

\end{document}